# Interfacial stress transfer in monolayer and few-layer MoS$_2$ nanosheets in model nanocomposites

Ming Dong[a], Robert J. Young[b], David J. Dunstan[a], and Dimitrios G. Papageorgiou[c*]

[a] School of Physical and Chemical Sciences, Queen Mary University of London, London E1 4NS, UK

[b] National Graphene Institute, Henry Royce Institute and Department of Materials, School of Natural Sciences, The University of Manchester, Manchester M13 9PL, UK

[c] School of Engineering and Materials Science, Queen Mary University of London, London, E1 4NS, UK

**Abstract**

Understanding the stress transfer mechanisms from a polymer matrix to two-dimensional (2D) reinforcements is essential for the preparation of high performance nanocomposites. In this study, the interfacial stress transfer from a flexible polymer substrate to monolayer and few-layer molybdenum disulfide (MoS$_2$) under tension has been investigated. Layer-dependent and strain-dependent photoluminescence (PL) spectroscopy were used to examine the stress transfer efficiency. The interlayer stress transfer efficiency of MoS$_2$ was determined to be in the range of 0.76–0.86, higher than that of graphene. The transfer of strain from the polymer substrate to the flakes was derived through strain-dependent band shifts. With progressive loading, the strain distribution in monolayer MoS$_2$ can be described by the shear-lag, partial-debonding and total-debonding models. The interfacial shear and frictional stresses were calculated to quantify the strength of the MoS$_2$/polymer interface. It was found that the strength of the interface is similar to the strength of a graphene-polymer interface. Strain mapping was performed at different strain levels and it was found that the strain distribution in bilayer MoS$_2$ is similar to the case of a monolayer sample. The interfacial shear strength remains almost unaffected, while the stress transfer length increases with increasing layer number.

*Corresponding author. email: d.papageorgiou@qmul.ac.uk

# 1. Introduction

Ever since the isolation of graphene [1], extensive research efforts have focused upon measuring the mechanical properties of graphene and other two-dimensional (2D) materials [2]. Graphene is considered to be the strongest material with a Young's modulus of 1 TPa and a tensile strength of 130 GPa [3]. The Young's modulus of hexagonal boron nitride (hBN) (0.86 TPa) is slightly lower than that of graphene, while the strength of hBN (70 GPa) does not degrade with the increase of layer number, up to nine layers [4]. Molybdenum disulphide ($MoS_2$) is another widely-researched 2D material, mainly due to its band gap that enables application in flexible optoelectronic, spintronic and valleytronic devices [5]. The in-plane stiffness of monolayer $MoS_2$ is comparable to that of steel (~270 GPa), while its fracture strength is around 23 GPa. [6]. The mechanical properties of 2D materials make them highly attractive as reinforcing agents for the fabrication of multifunctional polymer nanocomposites [7, 8].

When 2D materials are dispersed in a polymer matrix, strain is transferred to the nanofillers via interfacial shear. As the Young's modulus of the fillers is much higher than that of the matrix, higher stresses can be sustained and mechanical reinforcement can be achieved. Therefore, understanding the stress transfer mechanism between the filler and the matrix can help towards optimization of the mechanical reinforcement of polymer nanocomposites with 2D nanofillers. Raman spectroscopy has been widely used to monitor strain in 2D materials as the strain can induce shifts in the characteristic Raman bands of 2D materials [8-13]. Similarly, photoluminescence (PL) spectroscopy has also been used to characterize strain distribution in 2D materials with a direct band gap (such as transition metal dichalcogenides) as strain can modulate the band gap [14-16]. Both techniques allow precise monitoring of the strain transferred from a matrix to 2D materials and facilitate the study of interfacial and interlayer stress transfer mechanisms.

The stress transfer mechanism from a polymer matrix to graphene has been well studied [17, 18]. At low strains, the elastic stress is transferred by shear and the phenomenon can be described by the shear-lag theory, according to which, the stress builds up from the edges and reaches its maximum value towards the centre of the filler [17]. With the increase of applied strain, slippage initiates from the edges where the stress is transferred by friction and expands towards the centre of the flake [18]. Flakes with large lateral size (>30 μm) and interfaces with high shear strength are favoured to achieve high stresses in the flakes and hence effective reinforcement [17]. The effects of other factors, such as layer number [19, 20], edge effects [21, 22], wrinkles [23-25], chemical functionalization [26, 27], size effect [28] and cyclic loading [29], on the interfacial stress transfer have also been investigated for the case of graphene. There is a balance that needs to be considered during the design of graphene-based nanocomposites between the ability to achieve higher loadings by using thicker nanoplatelets and the reduction in effective Young's modulus of the reinforcement as layer number increases [19]. The stress transfer can deviate from the shear-lag theory for a distance of about 2 μm from the edges due to residual stress and doping [21]. Wrinkles with small amplitudes will not degrade interfacial stress transfer heavily as graphene still conforms to the substrate, while wrinkles of high amplitudes will reduce the interfacial stress transfer efficiency as delamination occurs [25]. In terms of chemical functionalisation, there is a competition between the reduced Young's modulus of the functionalised flakes and the increased interfacial shear strength [26]. All these efforts provide a critical insight into the interfacial stress transfer from a polymer matrix to graphene flakes and can offer guidance for the fabrication of high-performing graphene-reinforced polymer nanocomposites.

Studies on the fundamentals of stress transfer from a polymer to other 2D materials beyond graphene are quite limited [30-33]. Liu *et al*. [30] reported that the stress transfer efficiency highly depends on the Young's modulus of the matrix and only ~10% of applied strain can be transferred from a soft polydimethylsiloxane (PDMS) matrix to monolayer $MoS_2$. Wang *et al*. [31] found that interfacial stress transfer in monolayer $WS_2$ nanocomposites can be well described by the shear-lag theory. In another work, Wang *et al*. [32] concluded that multilayer hBN can be also considered as highly efficient load bearer as a result of its strong interlayer bonds. Recently, Liu *et al*. [33] reported that $Ti_3C_2T_x$ MXene with a flake length > 10 μm and a thickness of 10s of nanometers can offer efficient mechanical reinforcement to a polymer matrix. These studies promote the understanding of interfacial and interlayer stress transfer in 2D materials beyond graphene. However, the interfacial stress transfer from polymer to monolayer and few-layer $MoS_2$ has not been studied quantitatively. More specifically, key parameters governing interfacial stress transfer efficiency, such as interfacial shear stress, critical transfer length and interlayer stress transfer efficiency have not been determined yet. Considering the opportunities in using monolayer and multilayer $MoS_2$ to reinforce polymers (see [34-36] and references therein), it is crucial to study the deformation of $MoS_2$ within a polymer matrix and the underlying reinforcement mechanisms. In this study, the interfacial stress transfer from a polymer matrix to monolayer and few-layer $MoS_2$ flakes has been investigated using *in situ* PL spectroscopy under strain.

## 2. Experimental section

### 2.1 Materials

A natural 2H-$MoS_2$ crystal with a lateral size of ~1.5 cm was purchased from HQ Graphene, Groningen, Netherlands. The tape used for the mechanical exfoliation of $MoS_2$ flakes was supplied by Nitto Denko Corporation, Japan. The Gel-Film used for the transfer of the flakes was procured by Gel-Pak. A commercially available poly(methyl methacrylate) (PMMA) sheet was cut into beams with length of 70 mm and width of 20 mm by laser cutting. Strain gauges with grid resistance of 120 Ω were provided by VPG Corporate.

### 2.2 Mechanical exfoliation of $MoS_2$

$MoS_2$ monolayer and few-layer flakes were prepared by mechanical exfoliation of bulk $MoS_2$. The flakes were transferred onto a PMMA beam by an all-dry deterministic placement method [37]. Before transfer, the PMMA beam was washed with isopropyl alcohol, was then cleaned further in an ultrasonic bath and was rinsed with DI water. The process was repeated several times to ensure the cleanliness of the substrate and after that the beam was dried using nitrogen gas. For the model nanocomposite samples, a thin layer of PMMA (4 wt% solution in methoxybenzene, 2000 rpm for 1 min, ~ 200 nm thickness) was spin-coated on top of the flakes.

**2.3 Raman and PL spectroscopy**

Raman and PL spectra were collected using a micro-Raman spectrometer (Renishaw inVia) in a backscattering configuration. The laser beam was focused through a 100x objective lens and the laser power was kept below 0.3 mW to avoid laser heating during the experiments. For the primary characterisation of the layer number of $MoS_2$, PL spectroscopy was utilised using a laser excitation of 633 nm with a grating of 1200 grooves/mm and a laser excitation of 442 nm with a grating of 2400 grooves/mm. The integration time was 10 s. Raman spectroscopy was employed using a laser excitation of 442 nm with a grating of 2400 grooves/mm. The integration time was 30 s.

To deform the $MoS_2$ nanoflakes, the PMMA substrate with $MoS_2$ deposited on top of it (as illustrated in Figure 1a), was mounted on a four-point bending device. Strain was applied to the specimens in steps of about 0.05 % and the PL spectra were recorded from the middle of each flake. For line mapping measurements, the $MoS_2$ flakes were scanned from one end towards the centre of the specimens in steps of ~1 μm along the loading direction.

**3. Results and discussion**

**3.1 Deformation of $MoS_2$**

**3.1.1 Monolayer**

An optical image of a monolayer $MoS_2$ flake on the polymer substrate is shown in Figure 1b. The length of the flake along the loading direction was about 18 μm. The PL and Raman spectra of monolayer $MoS_2$ are shown in Figure S1. In the monolayer $MoS_2$, the indirect transition gap is larger than the direct transition gap, and the smallest band gap is therefore the direct band gap at the K point [38]. This attributes strong PL intensity to monolayer MoS$_2$, as shown in Figure S1a. As the layer number decreases, the Raman $A_{1g}$ mode softens due to weaker restoring forces in the vibrations, while the Raman $E^1_{2g}$ mode stiffens arising from either Coulomb interlayer forces or stacking-induced changes in the interlayer bonding [39]. For monolayer MoS$_2$, the frequency difference between $E^1_{2g}$ and $A_{1g}$ modes is the smallest (~ 19.2 cm$^{-1}$). These two features allow confirmation of the monolayer nature of MoS$_2$. The loading direction is also shown in Figure 1b (upper right) which is almost parallel to the long edge of the flake. The PL spectra of MoS$_2$ were collected at different strain levels. Considering the fact that the size of the flake is three orders of magnitude smaller than the length of the substrate, the flake was assumed to be strained uniaxially [9].

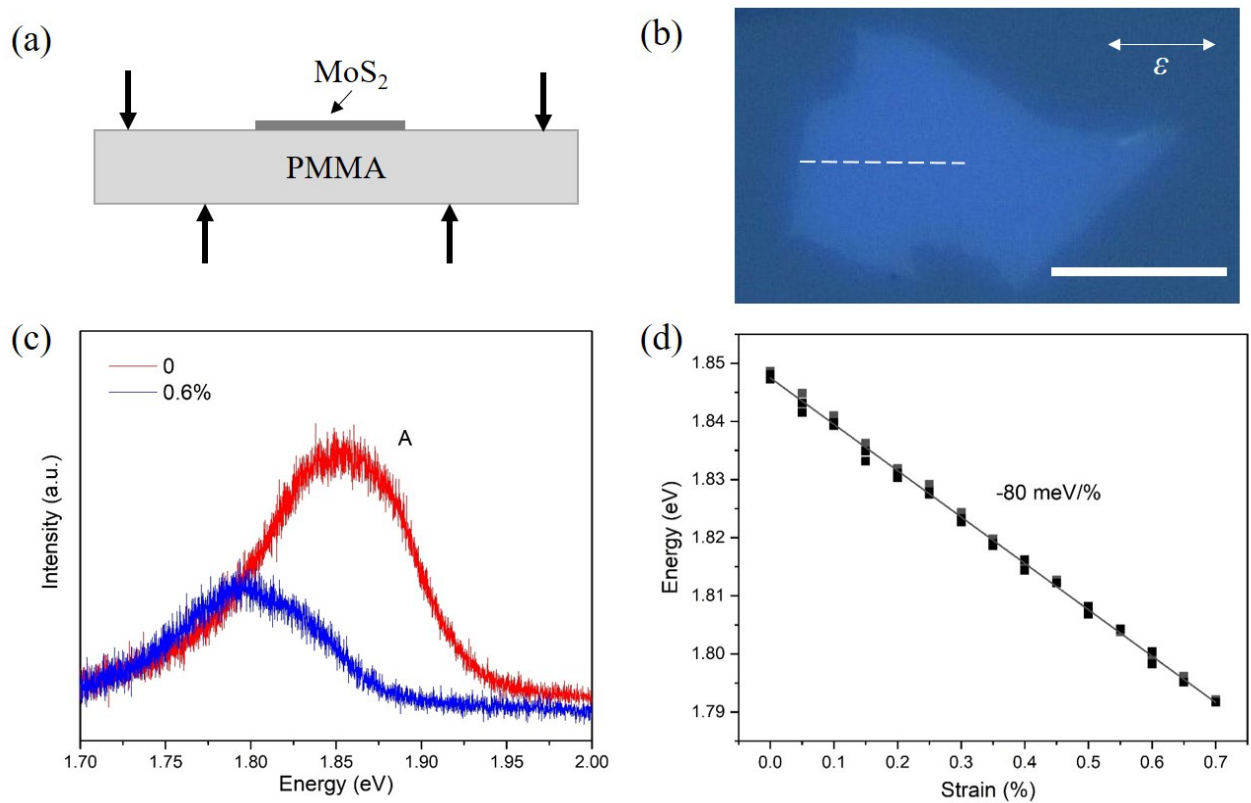

Figure 1. (a) Schematic diagram of the experimental setup, where MoS$_2$ flakes are supported on a PMMA substrate. (b) Optical image of a monolayer MoS$_2$ flake deposited on a PMMA substrate. Scale bar, 10 μm. (c) PL spectra of the monolayer flake under 0% and 0.6% strain, respectively. (d) Evolution of the A peak of the MoS$_2$ sample with applied strain from 0 to 0.7%.

The evolution of the PL peak measured at around the centre of the flake under strain is illustrated in Figure 1c and 1d. The principal PL peak (A peak, due to a direct band gap at the *K* point) was used as strain indicator because it has been shown that the A peak position provides a higher sensitivity to strain compared to Raman peak position [30]. The PL spectra were fitted with two peaks as shown in Figure S2a, where A denotes the exciton and T denotes the trion [40]. The A peak was fitted with a Gaussian function while the low intensity T peak was fitted with a Lorentzian function. The application of strain changes the PL spectra significantly. For the PL position, the A peak redshifted linearly with the increase of applied strain at a rate of −80 ± 1 meV/% (Figure 1d). The application of strain increases the Mo-Mo and Mo-S bond lengths and therefore reduces the orbital hybridization and d-bandwidth. This is reflected in the red-shifted exciton resonance energies and the reduced band gap [41]. It should be noted that the length of the exfoliated flake along the loading direction was about 18 μm, which for a monolayer flake on a PMMA substrate ($E \approx 3$ GPa) [30] is large enough to allow efficient stress transfer [17]. Therefore, the strain in Figure 1d is the strain in the flake, equal to the applied strain; this will be further demonstrated in Section 3.2.1.

**3.1.2 Interlayer stress transfer efficiency**

In order to study the effect of layer number on the stress transfer mechanism from the polymer matrix to a MoS$_2$ nanoinclusion, the deformation of monolayer, bilayer and trilayer MoS$_2$ was investigated. Before deformation, Raman and PL spectra were used to determine the layer number of MoS$_2$ flakes. The optical images and Raman spectra of MoS$_2$ flakes with different layer number are presented in Figure S3. It can be seen that the wavenumbers of $E_{2g}^1$ and A$_{1g}$ modes increase with the increase of layer number, in agreement with previous results [39]. The layer number can be further confirmed from the intensity of the PL peak as shown in Figure 2a. The PL intensity of monolayer

MoS$_2$ is strong due to the direct band gap and then decreases for bilayer and trilayer samples due to the transition from direct band gap to indirect band gap [38].

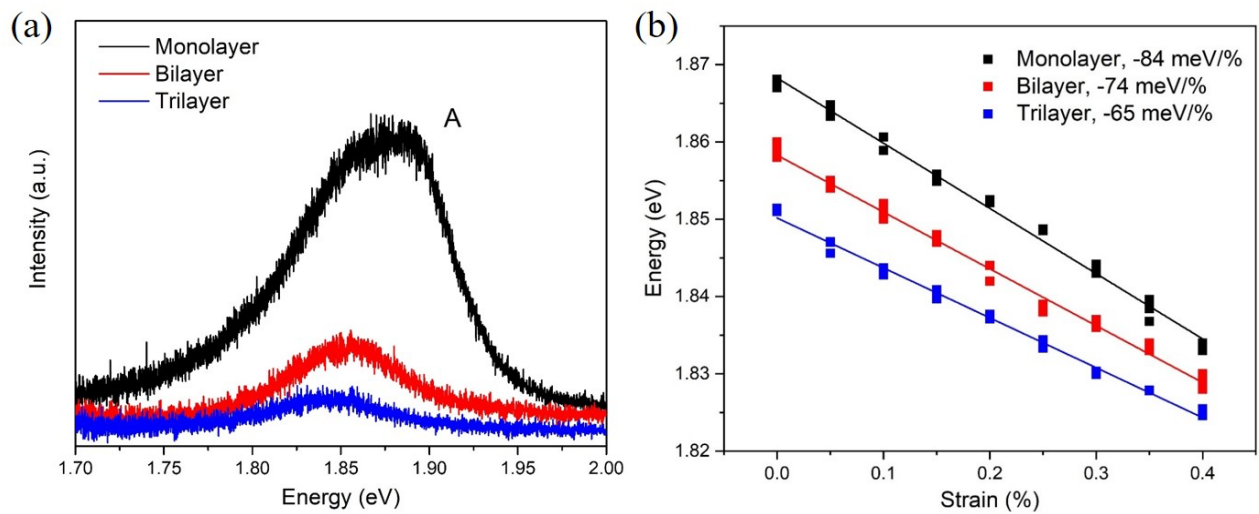

Figure 2. (a) PL spectra of monolayer, bilayer and trilayer MoS$_2$ at zero strain (b) Evolution of the PL A peak of monolayer, bilayer and trilayer MoS$_2$ with strain.

The redshifts of the characteristic PL A band for the monolayer, bilayer and trilayer MoS$_2$ flakes as a function of strain are illustrated in Figure 2b. The flakes are simply supported on PMMA substrates (there is no polymer coating). The strain-induced band shifts were recorded from the centres of the flakes. The slopes show a decreasing trend with the increase of layer number. For the monolayer sample, the redshift of the A peak with strain is the highest with a rate of −84 meV/%. This shift is slightly higher than the value of previous monolayer sample (−80 meV/%) (Figure 1c), an acceptable difference considering minor variations in the surface roughness of the substrate. The shift rate decreases to −74 meV/% for the bilayer sample, and −65 meV/% for the trilayer sample. Similar behaviour has been observed in graphene both simply supported and fully embedded in polymer substrates and was attributed to relatively weak van der Waals interlayer interactions and ineffective internal stress transfer [19].

Coated samples were also studied (where the MoS$_2$ flakes with different layer numbers were coated by a thin PMMA layer, mimicking a model nanocomposite) and the results are presented in Figure S4 and Table 1. The band shift rates of coated samples are higher than their uncoated counterparts. Within this "model" nanocomposite structure, stress can be transferred more effectively, while the top and bottom interfaces adhere to the polymer layers. The decrease in the band shift rate with the increase of layer number is also observed in coated samples. For coated monolayer, bilayer and trilayer samples, the band shift rates decreases from −92 meV/% to −86 meV/% and −76 meV/%, respectively, as shown in Figure S4b to S4d.

**Table 1.** Summary of PL band shifts for $MoS_2$ with different layer numbers. (F: 4 or 5 layers)

| Coating | Applied strain | Layers | PL A peak shift (meV/%) | Number of flakes |
|---|---|---|---|---|
| Uncoated | 0.4% | 1 | −82 ± 3 | 4 |
|  |  | 2 | −72 ± 1 | 3 |
|  |  | 3 | −64 ± 1 | 2 |
|  |  | F | −55 ± 2 | 2 |
| Coated | 0.4% | 1 | −92 ± 7 | 3 |
|  |  | 2 | −87 ± 2 | 3 |
|  |  | 3 | −76 ± 1 | 2 |
|  |  | F | −65 ± 4 | 4 |

To have a better understanding of the downward trend with layer number and avoid flake-to-flake variation, more flakes were deformed under similar experimental conditions. It should be noted that all the flakes studied here have relatively large lateral size (tens of microns) to ensure efficient strain transfer from the polymer matrix to the bottom layer of flakes. The strain applied was up to 0.4% and the spectra were taken from the centre of the flakes to avoid slippage at the $MoS_2$-polymer interface. From the results presented in Table 1 it can be seen that the flake-to-flake scatter is small and the downward trend is obvious. In previous studies, the band shift rates have been correlated to the efficiency of stress transfer from the polymer matrix to multiwalled carbon nanotubes [42] and multilayer graphene [19]. If the $MoS_2$-polymer interface remains intact, the band shift rate is an indication of stress transfer efficiency within the $MoS_2$ layers. The interlayer stress transfer efficiency, $k$ ($k = 1$ for perfect bonding and $k = 0$ for absence of bonding or complete slippage), can be related to the measured band shift rate, $(dE_g/d\varepsilon)$, through the following equation [19]

$$\left(dE_g/d\varepsilon\right)_{uncoated} = \frac{\left(dE_g/d\varepsilon\right)_{monolayer}}{[N - k(N - 1)]}, \tag{1}$$

where $(dE_g/d\varepsilon)_{monolayer}$ is the slope of monolayer $MoS_2$, and $N$ is the layer number. Using Equation 1 and the data in Table 1, the value of $k$ is calculated to be about 0.84–0.86 for uncoated samples. For coated samples, the above equation can be adapted to (for $N > 2$) [19]

$$\left(dE_g/d\varepsilon\right)_{coated} = \frac{\left(dE_g/d\varepsilon\right)_{monolayer}}{[N/2 - k((N/2) - 1)]} \tag{2}$$

Using the shift rates from coated values from Table 1, the interlayer stress transfer efficiency is calculated to be of the order of 0.76–0.84 for coated multilayer samples. These values are similar or slightly higher than the stress transfer efficiency in multilayer graphene (0.6−0.8) [19] as well as multi-walled carbon nanotubes (~0.7) [43], and lower than the value for multilayer hBN (~0.99) [32]. The difference in interlayer stress transfer efficiency originates from dissimilar interlayer shear strength and different sliding energies when subjected to strain.

The interlayer stress transfer efficiency can be used to determine the effective Young's modulus of the flakes [19, 44]. The effective Young's modulus, $E_{eff}$, can be determined by [19]

$$(E_{eff})_{uncoated} = \frac{E_{2D}}{[N - k(N - 1)]},$$

$$(E_{eff})_{coated} = \frac{E_{2D}}{[N/2 - k((N/2) - 1)]}, \tag{3}$$

where $E_{2D}$ is the Young's modulus of a 2D material. Using the average $k$ values (0.85 for uncoated and 0.8 for coated samples), it is calculated that the effective Young's modulus will be half of the $E_{eff}$ of the monolayer material when the layer number is higher than 7 and 12, for uncoated and coated samples, respectively. For graphene the $E_{eff}$ drops to half at a lower layer number (~6 for a coated sample) [19], while for hBN this value is extremely high (~100 for uncoated sample), indicating a very high interlayer shear strength [32].

## 3.2 Strain mapping of MoS$_2$

### 3.2.1 Monolayer

It has been demonstrated that stress transfer from a polymer matrix to a flake can be monitored by mapping strain along the flake [17]. Here, PL peak mapping was performed from the edge to the centre of the flake (dash line in Figure 1b) in steps of 1 μm, to gain an insight into the stress transfer from the polymer to monolayer MoS$_2$. It is necessary to convert the measured peak position to the real strain in MoS$_2$ flake; the band shift established in Figure 1c (−80 meV/%) was used in this calibration process. The strain distributions in the MoS$_2$ flake from the edge towards the centre of the flake at different strain levels are shown in Figure 3 and Figure S5. Considering the symmetry of the problem, half of the MoS$_2$/substrate system is illustrated here [22, 45]. Primarily, we need to evaluate the strain distribution in MoS$_2$ at low strain levels (Figure 3a). The strain distribution can be described by the conventional shear-lag theory with the assumption of a well-bonded polymer/2D material interface [17, 18]. This theory predicts the strain distribution in the MoS$_2$ flake, $\varepsilon_f$, as a function of the position, $x$, along the loading direction,

$$\varepsilon_f(x) = \varepsilon_m \left[1 - \frac{\cosh\left(ns\frac{x}{l}\right)}{\cosh(ns/2)}\right], \tag{4}$$

where

$$n = \sqrt{\frac{2G_m}{E_f}\left(\frac{t}{T}\right)}, \tag{5}$$

is the shear-lag parameter, $\varepsilon_m$ is the applied matrix strain, $G_m$ is the matrix shear modulus, $E_f$ is the Young's modulus of the flake, $l$ is the length in the $x$ direction, $t$ is the thickness of the flake, $T$ is the thickness of the representative volume and $s$ is the aspect ratio of the flake ($l/t$). The solid curve in Figure 3a is the fitting of Equation 4 to the experimental data at 0.28% applied strain with $ns = 10$. This $ns$ value is the same with the case of monolayer WS$_2$ flake supported by a PMMA beam [31]. The excellent agreement between the theoretical and experimental data indicate that the interface between the polymer and MoS$_2$ remains intact at such low strain and the strain distribution follows the expectation from shear-lag theory where the strain builds up from the edges towards the centre of the flake. The strain distribution allows the critical length $l_c$, which can provide indication on the quality of filler reinforcement and is small for strong interfaces, to be determined (twice the distance from the edge up to 90% of the maximum strain). At 0.28% applied strain, the critical length is estimated to be about 9 μm. This is similar with the $l_c$ of monolayer graphene simply supported on polymer substrate (that is 10 μm for 0.5% strain) [20], and larger than the $l_c$ of monolayer graphene embedded between a polymer substrate and a thin polymer layer in a "model nanocomposite" configuration (3 μm for 0.4% strain) [17]. This is reasonable as the stress can be transferred to monolayer and few-layer nanoplatelets more effectively when they are embedded into a matrix. For embedded flakes, more layers adhere to the surrounding polymer which means that more interfaces are stretched through interfacial shear. This can minimize the possibility of slippage during loading and optimize strain transfer efficiency.

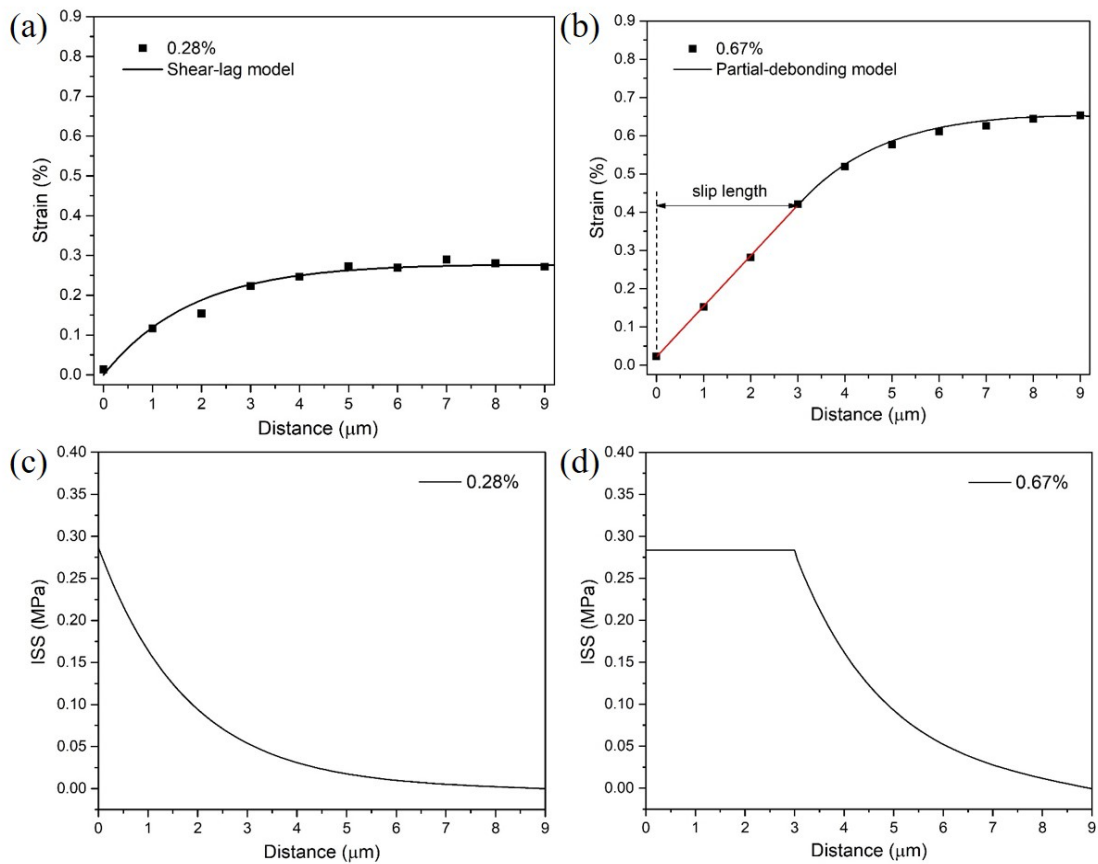

Figure 3. (a) Strain distribution in MoS$_2$ flake at 0.28% applied strain and the fitting result using the shear-lag model. (b) Strain distribution in MoS$_2$ flake at 0.67% applied strain and the fitting result using the partial-debonding model. The fitting parameter *ns* = 10 was used for both applied strains of 0.28% and 0.67%. (c) The interfacial shear stress distribution at 0.28% applied strain. (d) The interfacial shear stress and interfacial frictional stress distribution at 0.67% applied strain.

The variation of the interfacial shear stress (ISS), $\tau_i$, can be derived by [17]

$$\tau_i = nE_f\varepsilon_m \frac{\sinh\left(ns\frac{x}{l}\right)}{\cosh(ns/2)} \tag{6}$$

Using the fitting parameter *ns* = 10, the length set equal to 18 μm, the thickness of monolayer MoS$_2$ being 0.65 nm and the Young's modulus equal to 330 GPa [46], the shear stress distribution curve can be derived as shown in Figure 3c. It can be seen that the ISS is maximum at the edge and decreases to zero towards the centre of the flake. The maximum ISS is about 0.28 MPa. This value is similar with the one from graphene-polymer interfaces [17, 18, 20] as summarized in Table 2 and confirms the presence of van der Waals interactions between monolayer MoS$_2$ and the polymer matrix.

With the increase of strain, interfacial slippage will start from the ends and expand towards the centre of the flake. Similar behaviour has been observed in studies of other materials such as carbon fibres [47] and monolayer graphene [18]. In a previous study on graphene, the strain distribution after interfacial sliding was described by the nonlinear shear-lag model [18]. Unfortunately, by using this model, the strain in only the central part of graphene can be captured and the strain distribution near the edges (where slippage takes place) cannot be well-captured, as the shear-lag theory assumes no slippage. Herein, we show that the strain distribution in MoS$_2$ can be well-fitted by a partial-debonding model which has been used in the past to predict the strain distribution in carbon fibres under tension [47]. This model assumes that slippage occurs over a distance *ml*/2 (0 < *m* < 1) from the flake ends. In the sliding region, strain in transferred by friction and the strain distribution is almost linear as shown in Figure 3b and Figure S5a. In the centre of the flake, the interface remains intact and strain transfer follows the classical shear-lag behaviour:

$$\varepsilon_f(x) = \varepsilon_m - \left(\varepsilon_m - \frac{4\tau_f lm}{E_f t^2}\right) \frac{\cosh\left(ns\frac{x}{l}\right)}{\cosh\left[\frac{ns(1-m)}{2}\right]}. \quad (7)$$

In this case, the interfacial shear stress can be determined from the slope of the strain distribution using the force balance equation:

$$\frac{d\varepsilon}{dx} = -\frac{\tau_i}{ntE_f} \quad (8)$$

where $n$ is the number of flake layers. By introducing the measured slope of the slippage regions into Equation 8 (0.13%/μm at 0.67% applied strain), the interfacial frictional stress is estimated to be about 0.28 MPa as shown in Figure 3d. The frictional stress at 0.46% applied strain can also be estimated from Figure S5b (0.26 MPa), being similar with the value at 0.67% strain. The strain distributions at 0.46% and 0.67% were also fitted using the shear-lag model and the comparison between shear-lag model and partial-debonding model is shown in Figure 4. The shear-lag model overestimates the strain in the flake when slippage takes place. In contrast, the strain distribution is well-captured by the partial-debonding model, providing a more realistic evaluation of the interfacial shear stress at higher strains. Additionally, any errors originating from the incorrect use of the shear-lag model will increase with loading as a result of the slippage length further increasing with loading.

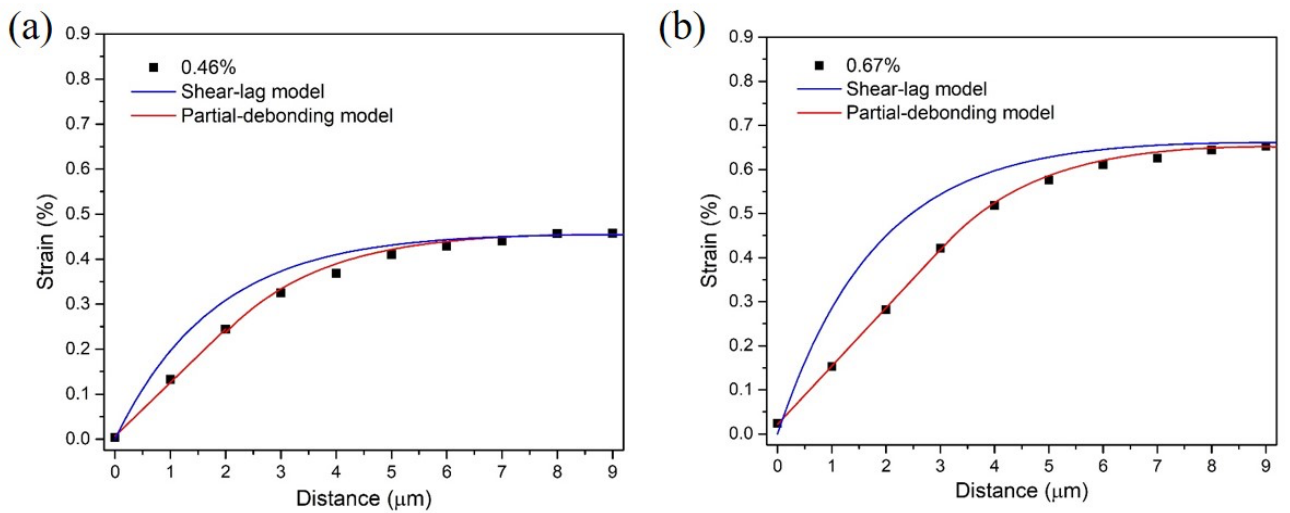

Figure 4. Strain distribution in monolayer MoS$_2$ at (a) 0.46% and (b) 0.67% applied strain. A comparison between the fitting curves of the shear-lag model and the partial-debonding model is presented ($ns$ = 10).

### 3.2.2 Bilayer MoS$_2$

Given the fact that polymers are rarely reinforced with monolayer flakes, the strain distribution in few-layer samples is also of high importance. Some key parameters such as the interfacial shear strength and transfer length can be determined through the study of strain distribution if there is no interlayer slippage [20]. In a previous study, it was shown that strain distribution is continuous between adjacent monolayer and bilayer areas within the same sample following the shear-lag behaviour for a coated bilayer graphene sample [19]. This is a clear indication that strain transfer from a polymer to monolayer and bilayer flakes follows similar rules. The optical image of an uncoated bilayer MoS$_2$ is shown in Figure S6a. The lateral size of the flake is about 32 μm which makes it long enough for efficient stress transfer. The PL A peak evolution with strain is shown in Figure S6b. The shift rate is about −71 meV/% which is close to the average value of bilayer samples shown in Table 1. This rate was used to convert the band gap shift to strain distribution in the bilayer flake. The lower conversion rate compared to the rate of monolayer MoS$_2$ (−80 meV/%) can originate from lower interlayer stress transfer efficiency (or possible interlayer slippage). Line mapping starts from the left edge towards the centre as shown in Figure S6a and the strain distributions at different applied strain levels are shown in Figure 5.

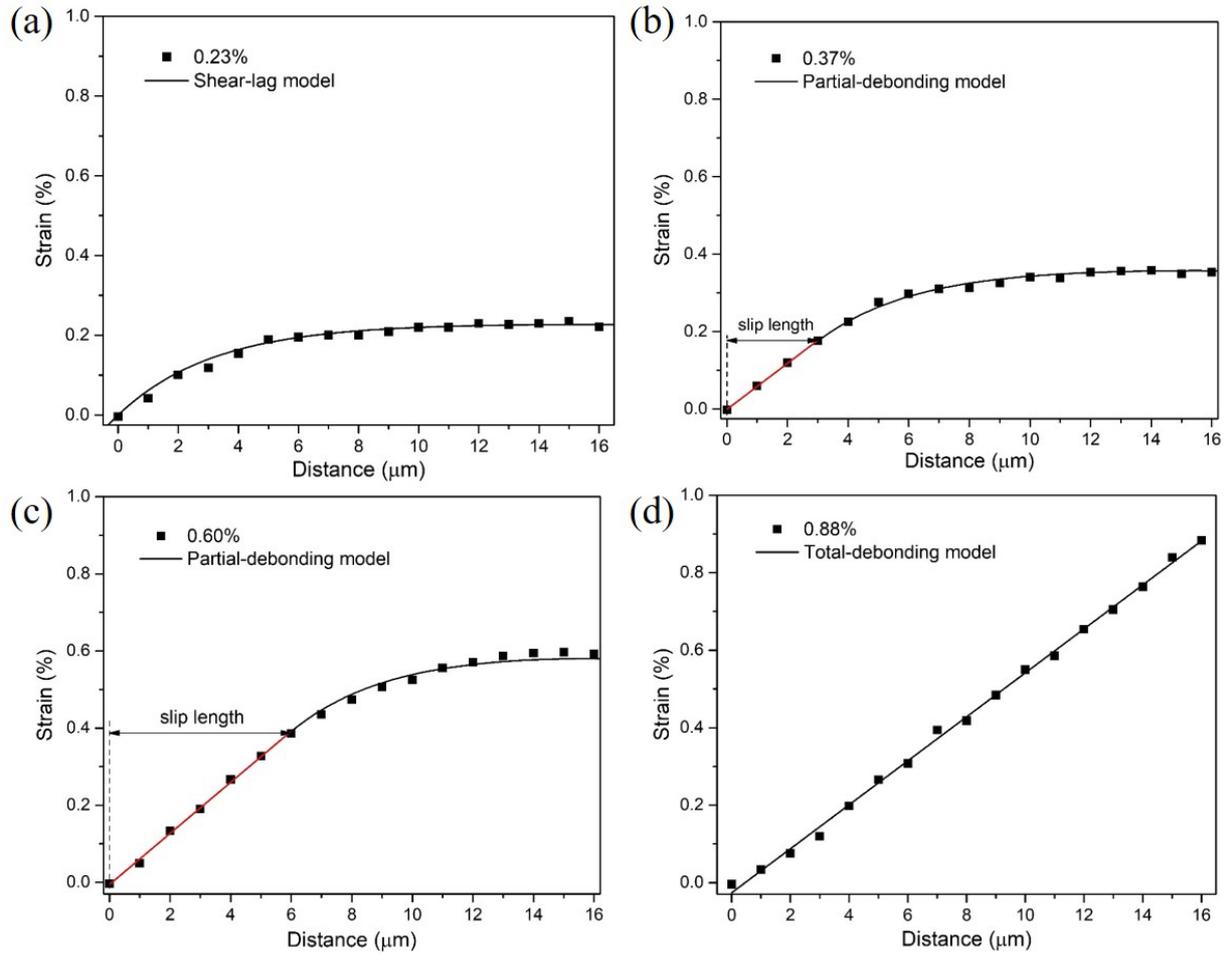

Figure 5. (a-d) Strain distributions in a bilayer, uncoated MoS$_2$ flake at different applied strain levels (0.23%, 0.37%, 0.60% and 0.88%). Shear-lag model, partial-debonding model and total-debonding model are used to fit the experimental results.

The evolution of strain distribution with applied strain is similar with the case of the monolayer sample (Figure 3). At 0.23% applied strain (Figure 5a), the strain distribution follows shear-lag behaviour. The fitting parameter *ns* = 10 is the same as the one used in the deformation of the monolayer flake (Figure 3a). The maximum ISS (as shown in Figure S7a) is observed at the edge of the flake with a value of 0.29 MPa, almost the same with the value of monolayer sample (0.28 MPa at 0.29% applied strain). This means that the bilayer MoS$_2$ flake is stretched as a whole and there is no interfacial slippage at such low strain. The interfacial stress transfer efficiency from the substrate to the monolayer sample and to the bottom layer of the bilayer sample remains at the same level. This confirms that the variations in the band shifts for samples with different layer numbers are the result of interlayer stress transfer efficiency. The transfer length is estimated to be about 14 μm for the bilayer flake, larger than the value for the monolayer sample. A similar variation has also been observed for multilayer graphene flakes [20]. This is explained by the fact that the applied strain is not fully transferred to the top layer of the bilayer sample due to relatively weak interlayer stress transfer efficiency ($k \approx 0.85$ for uncoated few-layer $MoS_2$).

With the increase of applied strain, slippage starts from the edge and expands towards the centre of the flake (Figure 5b and 5c). Similar to the case of the monolayer flake, the slip length increases with the increase of applied strain and can be quantitatively estimated by the partial-debonding model. For example, the slip length increased from about 3 μm at 0.37% applied strain to 6 μm at 0.60% applied strain. At 0.88% applied strain, the slip length expands to the whole flake and the stress is fully transferred by friction where the strain distribution can be fully captured by the total-debonding model [47]. The interfacial shear stresses derived from shear-lag behaviour and frictional forces are shown in Figure S7a to S7d. The maximum interfacial shear stresses are shown in Figure 6 for the monolayer and bilayer flakes to have a direct comparison. It can be seen that the variation in ISS is small for both monolayer (0.26–0.28 MPa) and bilayer (0.24–0.29 MPa) flakes. Similar behaviour has been observed in previous studies on graphene [20, 21]. This is not surprising as interfacial stress transfer takes place primarily through weak van der Waals interactions.

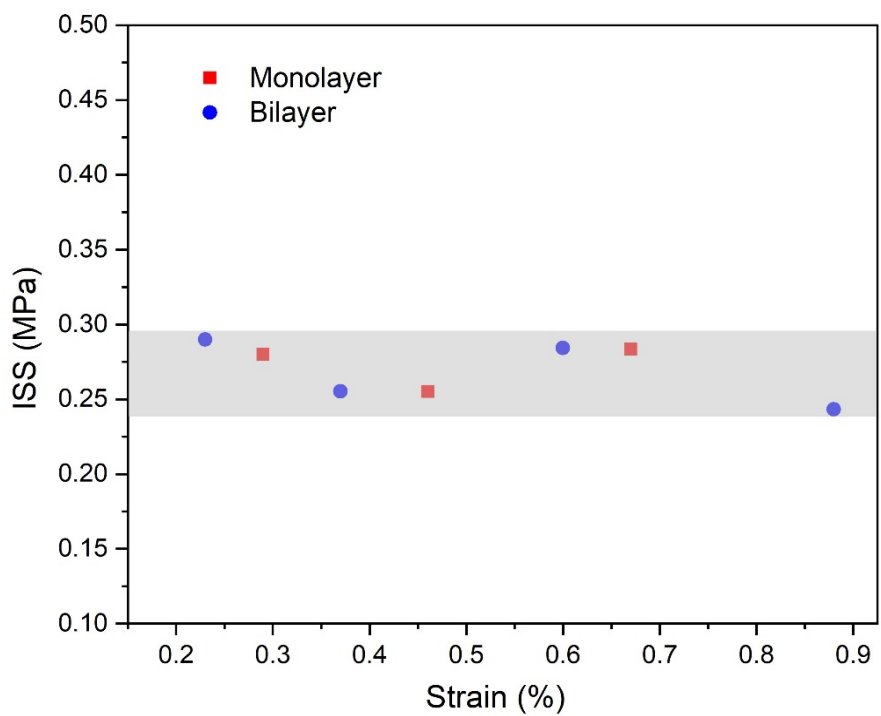

Figure 6. Variation of interfacial shear stress derived from shear-lag theory and force-balance equation for monolayer and bilayer $MoS_2$.

The parameters related to the interfacial stress transfer from a polymer matrix to various 2D materials are summarized in Table 2. Regardless of the polymer matrix (SU8, PET or PMMA), the interfacial shear stress for different 2D materials generally falls within the same order of magnitude. This means that for untreated interfaces the stress is transferred through van der Waals forces. There is a significant difference in the values of interlayer transfer efficiency originating from different interlayer shear strengths. For graphene, the interlayer stress transfer efficiency is the lowest due to the relatively low interlayer shear stress (about 40 kPa [48]). The interlayer stress transfer efficiency of $MoS_2$ lies between the values of graphene and hBN. Higher interlayer stress transfer efficiency means that the effective Young's modulus of the flakes will be less affected by the increase of layer number; this can have important implications on the implementation of few-layer $MoS_2$ in fields such as polymer nanocomposites given that monolayer materials are not expected to be used in bulk nanocomposites. This work has also revealed that for $MoS_2$ the transfer length required for efficient stress transfer will increase with layer number. Therefore, subsequent efforts towards production of MoS$_2$ nanoplatelets for effective mechanical reinforcement should primarily focus on the preparation of flakes with large lateral sizes.

Table 2. Parameters governing interfacial stress transfer from polymers to 2D materials.

| 2D materials | Layers | Polymer | Applied strain (%) | Interfacial shear stress (MPa) | Critical length (μm) | Interlayer transfer efficiency | Ref. |
|---|---|---|---|---|---|---|---|
| Graphene | 1 | SU8/PMMA | 0.4–0.6 | 0.3–2.3 | 3 | – | [17] |
| Graphene | 1 | PET | 1.2–7.0 | 0.46–0.69 | – | – | [18] |
| Graphene | 2 | SU8/PMMA | 0.4 | 0.15 | – | 0.6–0.8 | [19] |
| Graphene | 1 | SU8/PMMA | 1.6 | 0.2–0.5 | 10 | – | [20] |
| | 2 | SU8/PMMA | 1.5 | 0.2–0.5 | 15–22 | – | [20] |
| | 3 | SU8/PMMA | 1.6 | 0.2–0.5 | 22–30 | – | [20] |
| WS$_2$ | 1 | SU8/PMMA | 0.35–0.55 | 1.1–1.5 | 4 | – | [31] |
| Ti$_3$C$_2$T$_x$ | 1 | PMMA | 0.4 | 3–4 | – | – | [33] |
| hBN | 50 | PMMA | 0.1–0.3 | 3.8–9.4 | 6 | 0.99 | [32] |
| MoS$_2$ | 1 | PMMA | 0.7 | 0.26–0.28 | 9 | – | This work |
| | 2 | PMMA | 0.9 | 0.24–0.29 | 14 | 0.86 | This work |
| | 3 | PMMA | 0.4 | – | – | 0.76–0.86 | This work |
| | F | PMMA | 0.4 | – | – | 0.84 | This work |

## 4. Conclusions

In summary, the interfacial stress transfer from a polymer matrix to monolayer and few-layer MoS$_2$ flakes has been studied through the application of strain *in-situ* with photoluminescence spectroscopy. The strain evolution in monolayer and bilayer MoS$_2$ flakes follows the classic shear-lag theory at low strains and the partial-debonding and total-debonding models at higher strains. The key parameters governing the interfacial and interlayer stress transfer efficiency have been quantitatively determined. It has been demonstrated that the interfacial stress transfer efficiency between the MoS$_2$ layer and the matrix is similar with its graphene counterpart, while the interlayer stress transfer efficiency is higher than the one of graphene. Monolayer MoS$_2$ is more effective in terms of interfacial stress transfer compared to multilayer ones as the required transfer length is smaller. This study can further the understanding of interfacial stress transfer mechanisms from a polymer matrix to a large number of 2D nanoinclusions and can have important implications towards optimization of flake fabrication/exfoliation for the efficient mechanical reinforcement of bulk polymer nanocomposites.

**Acknowledgements**

M. Dong acknowledges support from the China Scholarship Council (CSC). D.G.P. acknowledges the support from "Graphene Core 3" GA: 881603 which is implemented under the EU-Horizon 2020 Research & Innovation Actions (RIA) and is financially supported by EC-financed parts of the Graphene Flagship.

# SUPPORTING INFORMATION

**Contents:**

**S1. Thickness identification of monolayer MoS$_2$**

**S2. PL peak fitting of monolayer MoS$_2$**

**S3. Raman spectra and optical images of monolayer and few layer MoS$_2$**

**S4. Band shifts of coated monolayer, bilayer and trilayer MoS$_2$**

**S5. Strain distribution in monolayer MoS$_2$**

**S6. Optical image, band shift and ISS of a bilayer MoS$_2$ flake**

**S7. Interfacial shear stress in a bilayer MoS$_2$**

## S1. Thickness identification of monolayer $MoS_2$

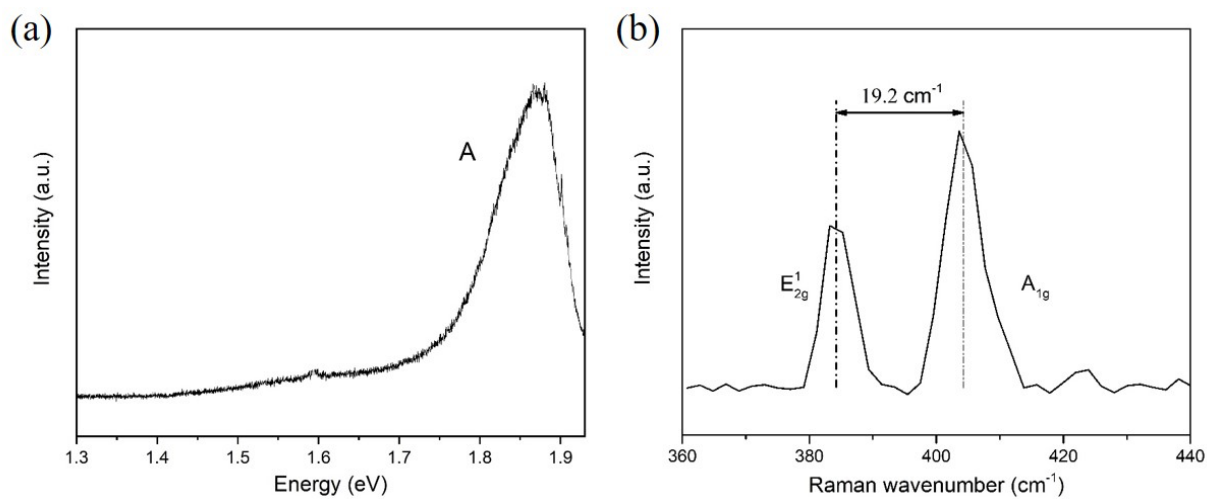

Figure S1. (a) PL spectra and (b) Raman spectra of monolayer $MoS_2$ on a PMMA substrate. The laser lines of 633 nm and 442 nm were used, respectively.

## S2. PL peak fitting of monolayer $MoS_2$

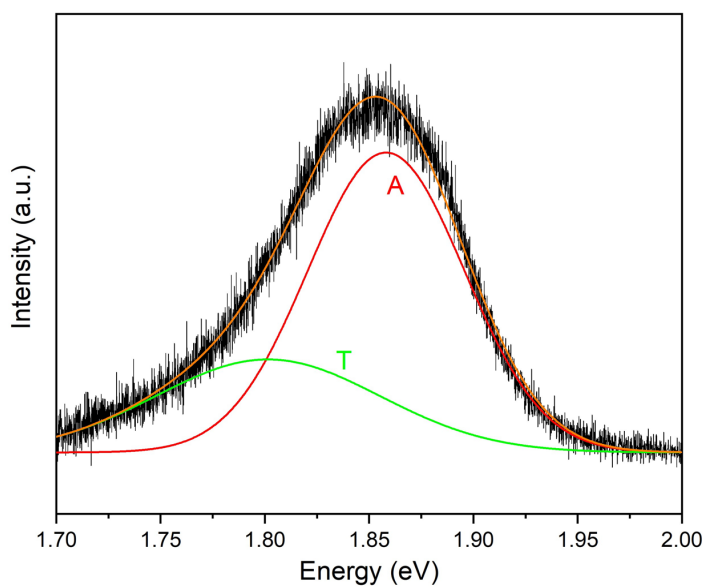

Figure S2. (a) PL spectrum of a monolayer $MoS_2$. The peak was fitted (orange curve) using two peaks, where A denotes the exciton (red curve - fitted with a Gaussian peak) and T denotes the trion (green curve fitted with a Lorenzian peak).

**S3. Raman spectra and optical images of monolayer, bilayer and trilayer MoS$_2$**

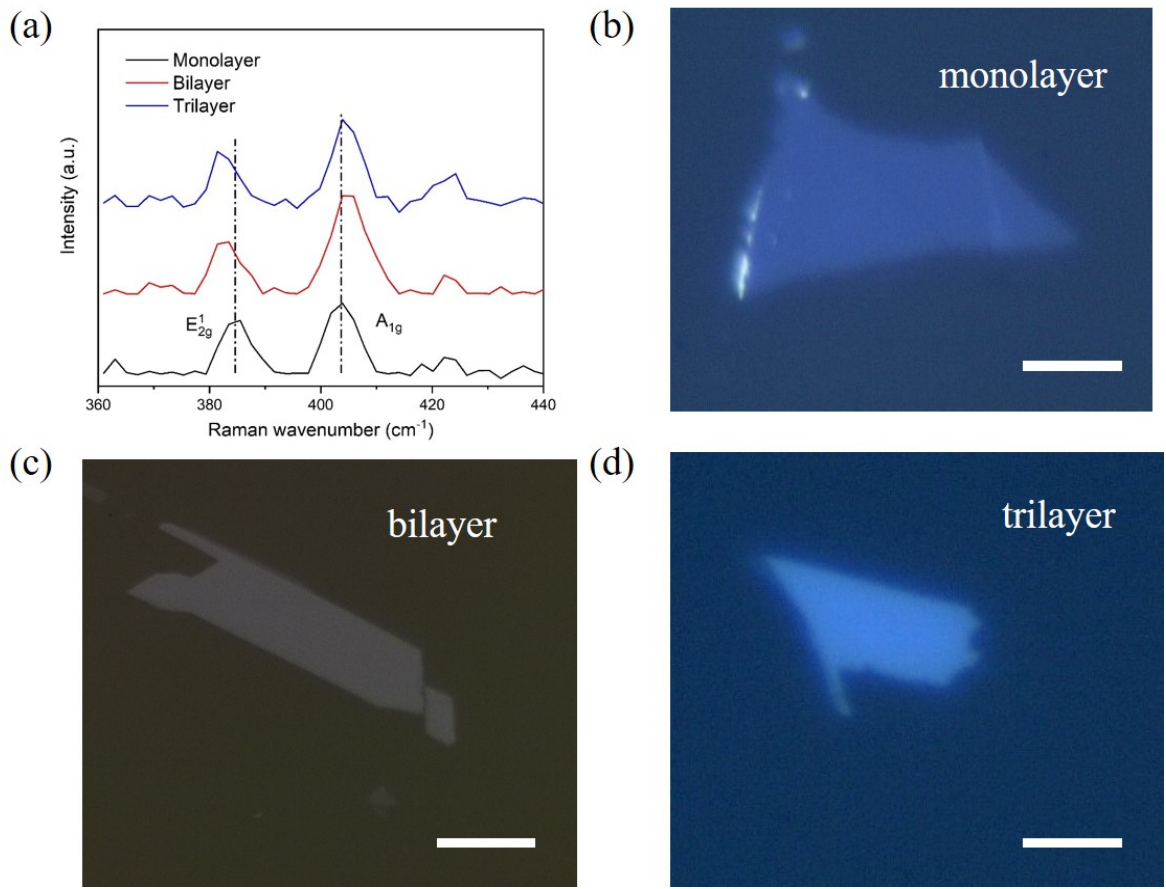

Figure S3. (a) Raman spectra of monolayer, bilayer and trilayer MoS$_2$. (b-d) Optical images of monolayer, bilayer and trilayer MoS$_2$, respectively. Scale bars, 10 μm.

## S4. Band energy shifts of coated monolayer and few layer MoS$_2$

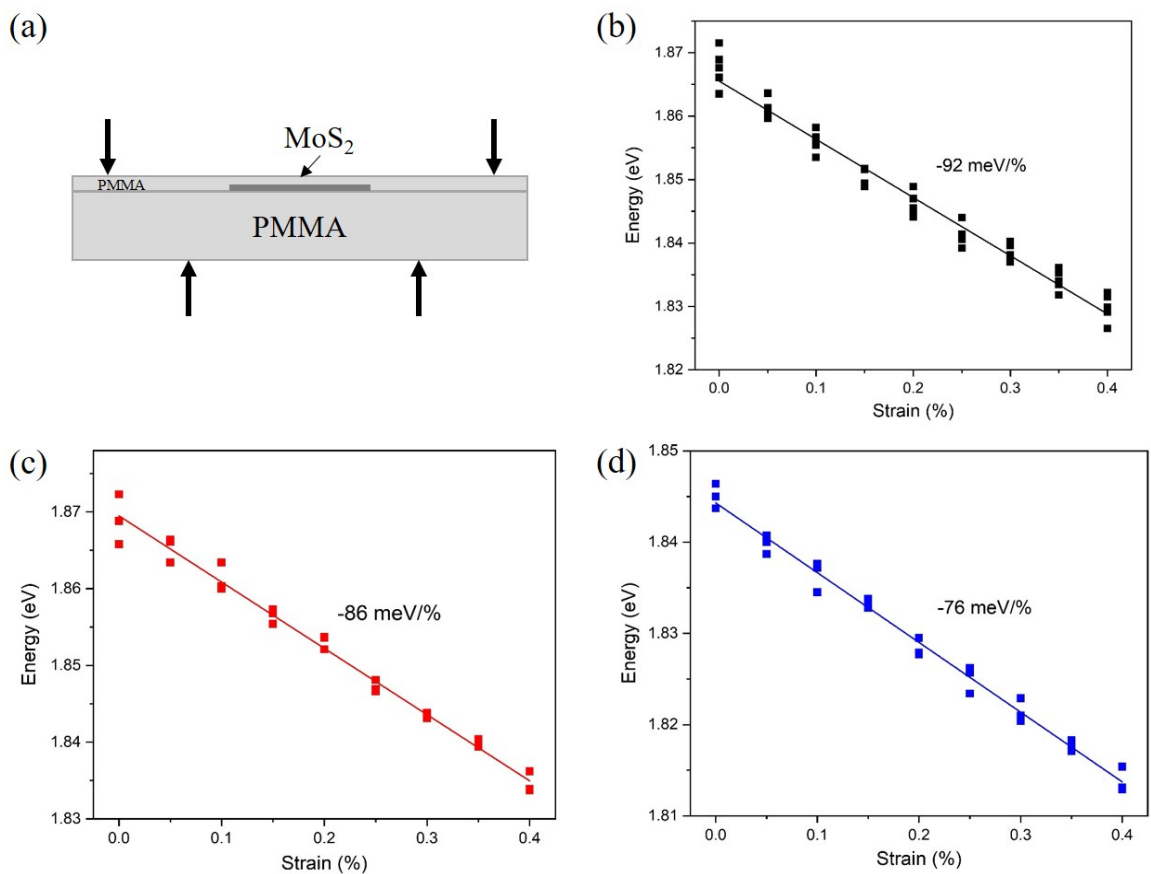

Figure S4. (a) Scheme of MoS$_2$ flakes embedded in PMMA layers. (b-d) Band energy shift of coated monolayer, bilayer and trilayer MoS$_2$, respectively.

## S5. Strain distribution in monolayer MoS$_2$

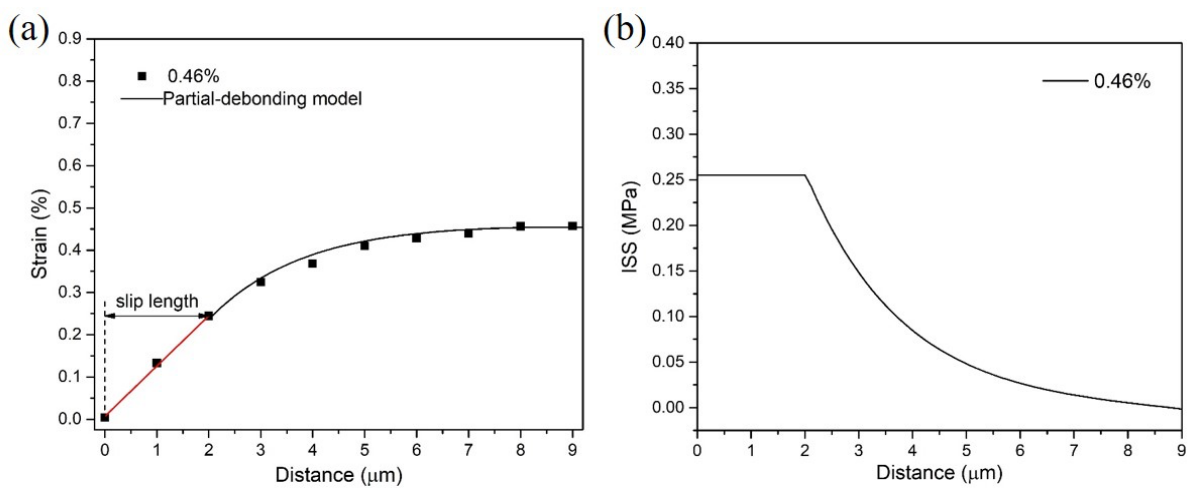

Figure S5. (a) Strain distribution in MoS$_2$ flake at 0.46% applied strain and the fitting result using the partial-debonding model. The fitting parameter *ns* = 10 for the applied strain. (b) The interfacial shear stress and interfacial frictional stress distribution at 0.46% applied strain.

## S6. Optical image, band gap shift, and ISS of a bilayer MoS$_2$ flake

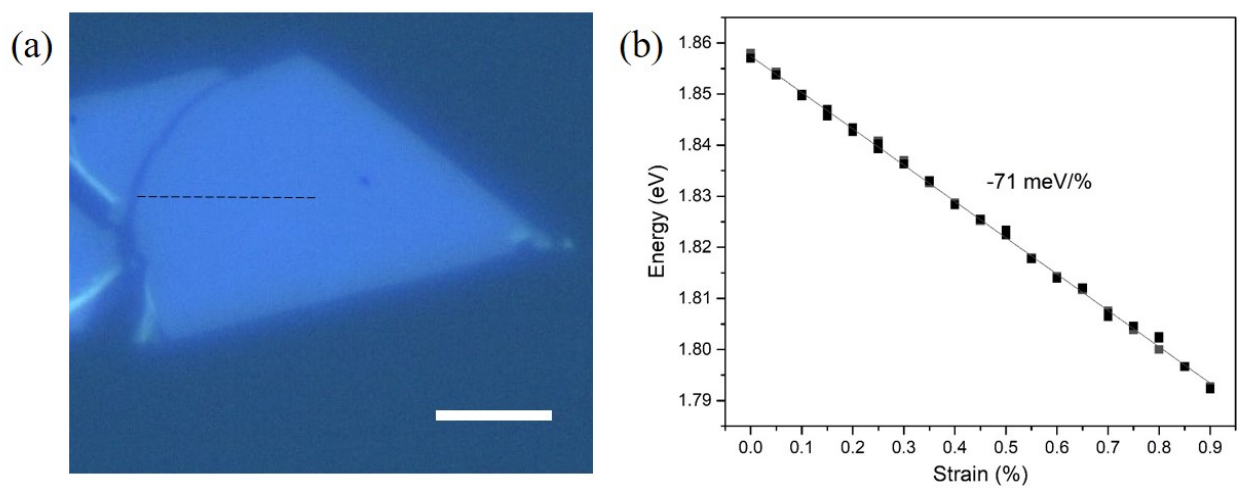

Figure S6. (a) Optical image of a bilayer MoS$_2$ flake. Scale bar, 10 μm. (b) The evolution of PL A peak with strain up to 0.9%. The A peak shift rate is about −71 meV/%.

**S7. Interfacial shear stress in bilayer MoS$_2$**

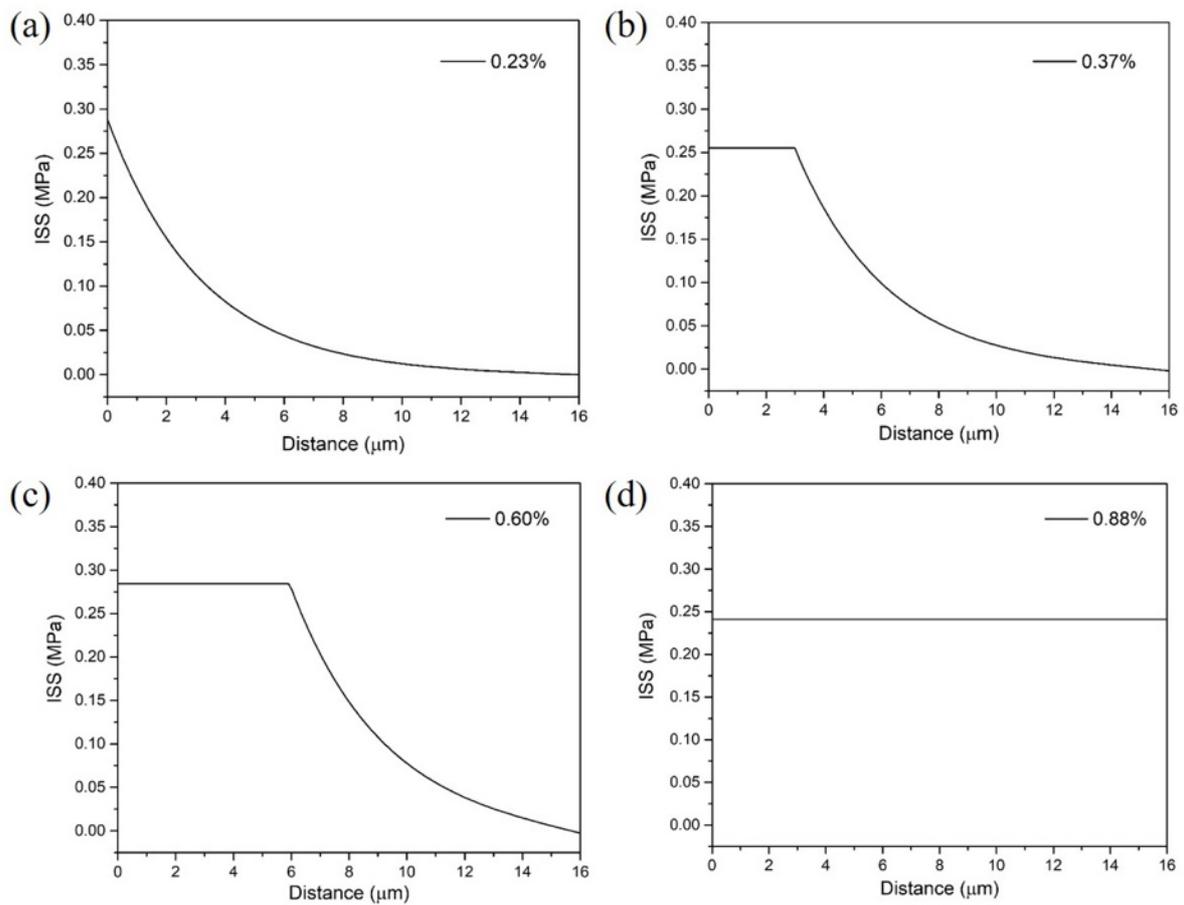

Figure S7. (a-d) Interfacial shear stress distribution of the bilayer MoS$_2$ sample at strain of 0.23%, 0.37%, 0.60% and 0.88%, respectively.